# Ethernet Topology Discovery: A Survey

Kamal A. Ahmat
CITY UNIVERSITY OF NEW YORK/Information Technology
New York, USA

Email: kamal.ahmat@live.lagcc.cuny.edu

**ABSTRACT**

**Ethernet networks have undergone impressive growth since the past few decades. This growth can be appreciated in terms of the equipment, such as switches and links, that have been added, as well as in the number of users that it supports. In parallel to this expansion, over the past decade the networking research community has shown a growing interest in discovering and analyzing the Ethernet topology. Research in this area has concentrated on the theoretical analysis of Ethernet topology as well as developing tools and methods for mapping the network layout. These efforts have brought us to a crucial juncture for Ethernet topology measurement infrastructures: while, previously, these were both small (in terms of number of measurement points), we are starting to see the deployment of large-scale distributed systems composed of hundreds or thousands of monitors. As we look forward to this next generation of systems, we take stock of what has been achieved so far. In this survey, we discuss past and current mechanisms for discovering the Ethernet topology from theoretical and practical prospective. In addition to discovery techniques, we provide insights into some of the well known open issues related to Ethernet topology discovery.**

## I. INTRODUCTION

This survey focuses on measurements of the *Ethernet network topology*, i.e., the representation of the interconnection between directly connected peers in the Ethernet network.

While information about network devices (i.e nodes) and connections can be obtained by processing the data collected from the network and passive measurements, researchers largely obtain information about network nodes, topology and its characteristics from active measurements.

There are three different levels at which to describe the network topology: the *link layer topology*, the *network layer topology*, sometimes referred to generically as the *internet topology*, and the *overlay topology*. The Internet topology can itself be seen at four different levels. The first one, the *IP interface level*, considers IP interfaces of routers and end-systems. Usually, this topology is obtained by using data collected with a probing tool such as *traceroute* [14]. The second level, *the router level*, treats each router as a single node in the topology graph.
It can be obtained by aggregating IP interfaces through a technique called *alias resolution* [19, 22, 29, 35]. The point of presence (PoP) level, is a third level, that can be obtained by further aggregating the routers, or directly aggregating the interfaces, that are identified as being geographically co-located.

Finally, the *AS level* provides information about the connectivity of autonomous systems (ASes). This information is not primarily drawn from active measurements, but rather from inter-domain routing information and address databases. However, a deep description of the Internet topology discovery mechanisms is beyond the scope of this article.

A typical overlay topology would be the topology of a peer-to-peer system. An overlay topology can be *unstructured* or *structured*. Structured overlays are exemplified by distributed hash tables, such as Chord [30] or CAN [24]. As explained by Stutzbach *et al.* [31]," peers select neighbors through a predominantly random process. An overlay topology is influenced by peer participation (i.e., join and leave mechanisms) as well as the protocol behavior (i.e., neighbor selection). Characterizing an overlay topology can be done by examining properties of snapshots of the overlay." These snapshots can be gathered using a topology crawler, an engine that queries peers for a list of their neighbors [31-32].

As stated by Stutzback *et al.* [32],"a deep understanding of the topological characteristics in overlay systems is required to meaningfully simulate and evaluate the actual performance of the proposed search and replication techniques."

The overlay topology has drawn the attention of the net- in this article; we are not directly concerned with peer-to-peer systems. Consequently, describing the overlay topology in more detail would be beyond the scope of this survey. Interested readers might refer to the work of Ripeanu *et al.* [26], Stutzbach *et al.* [32] and Liang *et al.* [21].

The link layer topology, the subject of this article, as defined by Breitbart *et al.* [5], refers to the characterization of the physical connectivity relationships that exist among entities in a communications network. In other words, it is the description of how data link layer devices, switches and bridges, are interconnected and how the different hosts are connected to them. Figure 1 depicts a simple typical Ethernet network.

Maintaining an accurate and complete knowledge of the link layer topology is a prerequisite to many critical network management tasks such as network diagnostics and resource management.

There is considerable scientific literature devoted to techniques for the discovery of link-layer topology. This research was mainly led by Breitbart *et al.* [5-7], Lowekamp *et al.* [22], Black *et al.* [2], Bejerano [3, 4], and, more recently, Gobjuka *et al.* [15-18].

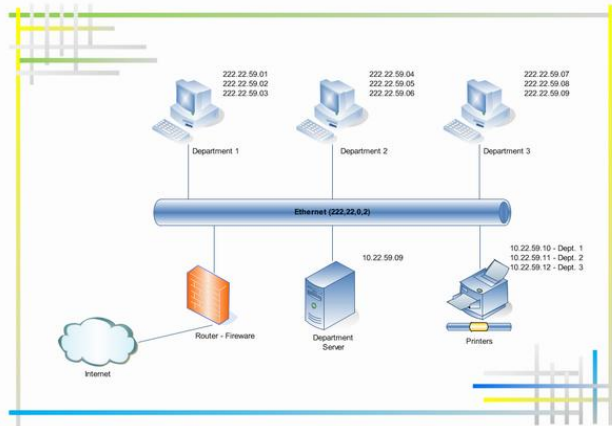

**Figure 1:** Simple Ethernet network.

The rest of the paper is organized as follows. The next section describes the motivation behind discovering topology of Ethernet networks. Section III describes methods used to discover Layer-2 network elements. In Section IV we focus on topology discovery methods presented in the literature. Section V describes limitations and issues related to Data Link topology discovery. Finally, Section VI concludes the paper.

## II. MOTIVATIONS

Network topology information can be valuable in a variety of situations; it can be used for network administration (including fault-detecting and avoiding [2, 5], network inventory and planning [5, 6], protocol and routing algorithm development [11], performance prediction [22] and monitoring as well as accurate network simulation [20]. From a network security perspective, topology information can find application in threat detection [1], network monitoring [37], network access control [10] and forensic investigations [12, 25].

Manual network mapping is becoming increasingly difficult [9] (if not impossible [38]) due to the size and dynamic behavior of networks. Automatic topology discovery tools and algorithms will therefore play an important role in network security, management and administration.

Research efforts concerned with physical topology discovery have focused mainly on cooperative network environments [2] where it is assumed that network elements are intelligent and can be queried for topology related information.

### A. Administration and Planning

Network administrators are often faced with network problems where fault-detecting and avoiding need to be performed [2, 39]. In order to troubleshoot network problems, a topology map of the network can effectively be used to isolate the problem area [4, 46]. The topology map can also help identify infrastructural vulnerabilities and the network can then be adapted to provide more redundancy.

From network management prospective, network topology information can be applied to network management. Network topology information is useful in deciding where to add new routers and to figure out whether current hardware is correctly configured. It also allows network managers to find bottlenecks and failures in the network.

Also, network expansion planning and decisions regarding the placement of new infrastructure are also aided by accurate knowledge of the network topology.

Network Management Systems (NMS) also employ topology information to help with network administration. The most notable systems include IBM's Tivoli1, Hewlett-Packard's OpenView2 and the open source Open-NMS3.

### B. Performance Prediction

In a second application area, that of performance prediction, topology information can be used to optimize the performance of network aware applications as well as the performance of distributed, either grid or cluster, applications.

Topology knowledge can help determine if a given network would provide a certain Quality of Service (QoS). As an example, in order to determine if a network would support multimedia technologies such as Voice over IP (VOIP), knowledge of the network topology is essential.

Multimedia content is increasingly shared between Ethernet network users. In order to improve the quality of service (QoS) offered to users and provide a high availability of the shared data, it is common to store the data in replicated servers distributed across the internet. The replication of data over different machines makes the choice of its location a challenging problem that can be addressed with knowledge of the internet topology.

### C. Algorithm and Protocol Design

Protocol design can use network topology knowledge. For instance, Radoslavov *et al.* discuss the impact of topology on the design and evaluation of four multicast protocols [40]. Also, a network's topology influences the dynamics of routing protocols [40] and should therefore be taken into account during the design of the protocols [24].

Large network topology visualization has proved to be a challenging task and algorithms have been developed for effectively presenting the topology information [24].

### D. Simulation

The accuracy of network simulations, a fourth application area, depends on realistic and accurate network topologies [31]. Generated topologies for use in simulation do not always match real-world topologies [32] and create the need for accurately measuring real-world network topologies.

Network simulation can not only help researchers understand the current behavior of a network, but also the effects of possible future changes to the network.

*E. Security*

Knowledge of the internet topology might have some applications in security. For instance, Burch and Cheswick propose to use internet topology information to track anonymous packets back to their source [9].

Firewalls have traditionally been placed at the network edge to protect against external threats. Insider threats to networks have become more common and it is estimated that they account for around 30% of security incidents. These security incidents also lead to significant financial losses [10]. Firewall placement and the management of a network security policy should therefore be influenced by the network topology.

Another perimeter defense mechanism, Intrusion Detection Systems (IDS), can also benefit by taking network topology information into account. If an IDS is not placed correctly it could generate both false positives and false negatives.

The problems with firewall and intrusion detection systems have generated research interest in the areas of Network Access Control (NAC, also called Network Admission Control [10]). These systems are proactive and attempt to enforce a network security policy at either layer 2 or 3 of the network [10].

Devices that do not conform to the security policy can for example be denied access to the network infrastructure by physically disabling switch ports.

A lack of knowledge about the network's topology and the connected devices can however seriously hamper the effectiveness of NAC solutions [10].

## III. NETWORK NODE DISCOVERY

The first step in gaining knowledge about an Ethernet network is identifying unique network nodes. Ethernet network nodes can be active, or passive. While the first type of nodes (e.g. switches) can be used to obtain information that can be used in the node and topology discovery process, passive nodes (e.g. hubs) don't provide any useful information that can be used for the discovery process.

Nodes in an Ethernet network are uniquely identified by their MAC addresses at layer 2, but this raw number by itself does not provide a lot of information about the node. Other sources of information were therefore combined, where possible, with this number to provide more information about each node.

Even though dumb network devices, such as hubs, may be transparent to the network, influence the performance and behavior of the network. Thus, it is significantly important to discover the presence of such nodes and their accurate locations and interconnectivity with other *visible* devices.

## IV. NETWORK TOPOLOGY INFERENCE

A network's physical topology can potentially correspond to several logical topologies depending on the level of abstraction used. In 2000 Breitbart *et al* [5] realized that network management tools as well as previous research efforts focused on layer 3 topology discovery and ignored the connectivity of layer 2 network elements. Where layer 2 topology discovery tools did exist, they were found to specifically target single vendor products [5]. Breitbart *et al* therefore developed algorithms that could perform layer 2 topology discovery in multi-vendor (heterogeneous) networks (See Figure 2) by using standard Simple Network Management Protocol (SNMP) Management Information Base (MIB) data.

The initial algorithm developed by Breitbart *et al* [5] depended on *perfect* Address Forwarding Table (AFT) data collected from every single element in the network.

Breitbart *et al* also observed that for multisubnet networks the network topology may not be unique even for the set of complete AFTs obtained from a simple Ethernet network. Breitbart *et al* proposed a newer algorithm that could successfully discover the target network topology, provided that the network was uniquely described by the SNMP MIB data obtained [6].

In such a case finding an exact topology is not possible. However, their algorithm generates some network fragments that can be uniquely determined.

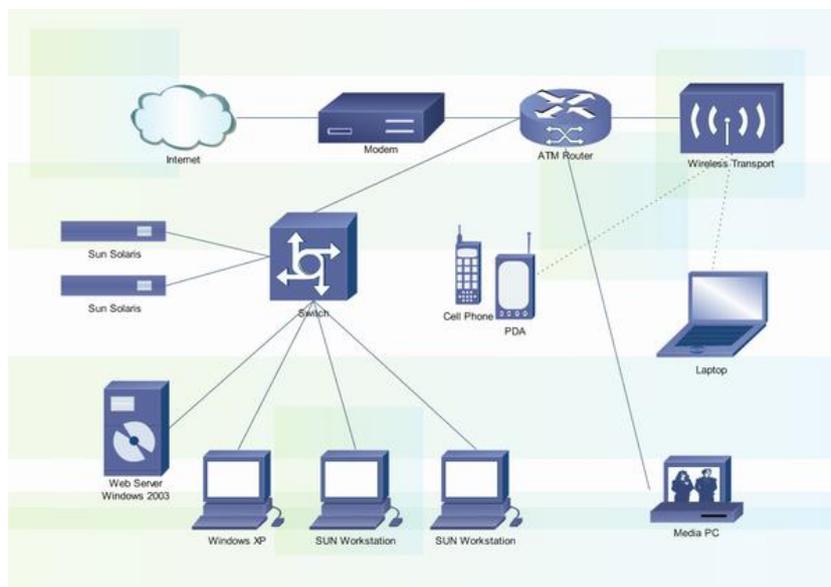

**Figure 2:** A typical multi-vendor, multi-protocol based network.

Lowekamp *et al.* relaxed the dependency on complete AFTs information [22] and proposed a necessary and sufficient condition for two AFTs to be connected (directly or indirectly). Their work also addressed the topologies that may contain uncooperative nodes, which are the nodes that can appear in other nodes' AFTs but don't provide access to their own AFTs. The work described in [22] could discover the topology with only limited AFT data collected from SNMP enabled network elements. However, their approach may fail to discover the topology even in simple networks as observed by Gobjuka and Breitbart [16, 17].

Bejerano *et al.* [3] proposed the first formal algorithm to discover the topology in presence of uncooperative elements (i.e. hubs.) Uncooperative elements do not speak SNMP, do not allow access or do not even have layer 2 addresses. The main issue with this algorithm was its complexity; the algorithm was too complex to understand and implement in practice. Furthermore, this method may not discover any topology if the given input set of AFTs defines a non-unique topology.

Sun et al [33, 34] proposed an algorithm based on "*connections reasoning technique*" that was claimed to be necessary and sufficient to discover the layer 2 topology even when the information provided by nodes MIBs is incomplete. However, their claim was not supported by proofs. Furthermore, the incorrectness of these claims was shown by Gobjuka and Breitbart [16, 17] by proving that discovering Ethernet topology when AFTs are incomplete is, in fact, an NP-hard problem, even if the network comprises a single subnet.

Further work by Bejerano [4] showed the limitations of the algorithms developed by Lowekamp *et al* [22] and Breitbart *et al* [5, 6] in multi-subnet networks or in the presence of uncooperative switches and hubs. Bejerano's algorithm was simple and could discover the topology in most of cases. However, it cannot guarantee a topology discovery. Also, his method also requires a completeness of input AFTs.

Research by Stott [41] also employed SNMP MIB data, but instead of using forwarding table data, the algorithm used data from the Bridge-MIB. However, the method described in this paper assumes that each device has knowledge of the spanning tree root, which doesn't happen always in practice.

Gobjuka and Breitbart [7, 15] described the first formal method to determine whether a given set of complete AFTs define a unique topology when the network doesn't contain hubs. They also showed that there is proportional relationship between the number of subnets in the network and non-uniqueness of the discovered topology.

Further work by Gobjuka and Breitbart [18] described the first practical algorithm to discover the Ethernet network topology when the network contains hubs. Their methods discover the all network topologies when the MIB information defines more than one topology. Furthermore, they proposed criteria to decide the uniqueness of network topology from a complete set of AFTs when the network contains hubs.

More recently, Gobjuka and Breitbart [16, 17] investigated the problem of finding the layer 2 topology for networks that may include uncooperative nodes when the available AFTs are incomplete. They proved that finding a layer 2 network topology for a given set of incomplete AFTs is an NP-hard problem even for single subnet networks and deciding whether a given set of AFTs defines a unique network topology is a co-NP-hard problem. The authors showed that the topology discovery problem is NP-hard even if there are two nodes "*a*" and "*b*" in the network such that node "*a*" appears in some AFTs and node "*b*" appears in some AFTs but neither "*a*" nor "*b*" appears in all AFTs. This condition was probably the strongest which makes the problem NP-hard as they also showed that the topology can be discovered in polynomial time if all AFTs include node "*a*". They also proposed heuristic algorithms to find network topology [16, 17]. Their also described methods for inferring complete AFTs from incomplete information. This approach is used in heuristic that discovers the topology from incomplete AFTs.

The Internet Engineering Task Force (IETF) attempted to create a standard for SNMP topology discovery by creating the Physical Topology MIB, but adoption of the proposal was hampered by the fact that it did not include details on how to actually populate the required MIB objects. To remedy the situation, the IEEE developed the Link Layer Discovery Protocol (LLDP) as part of the 802.1AB-2005 standard. The LLDP allows neighboring devices to become aware of each other and populate their Physical Topology MIBs. The efforts surrounding LLDP clearly shows an industry need for topology discovery in heterogeneous networks at layer 2; however, LLDP cannot easily be deployed on legacy equipment.

All the layer 2 topology discovery techniques and algorithms discussed thus far depend on SNMP enabled network elements. The reliance on SNMP can prove problematic in quite a number of network environments. As networks grow and management becomes decentralized it cannot be assumed that SNMP would be enabled or that administrative SNMP access would be granted. A lot of small business, home office and branch office networks are built using consumer-grade network equipment that do not even support SNMP.

A need for topology discovery techniques that do not require network cooperation and for tools that can augment SNMP-based techniques therefore exists.

A technique for layer 2 topology discovery without network element cooperation has been implemented by Black et al [2]. The technique exploits the packet forwarding properties of network elements, specifically those of switches. The algorithm requires specialized software on many edge nodes (hosts) that are controlled from a master node to execute the distributed discovery algorithm [2]. Cooperating hosts train switches they are connected to in order to only pass packets with specific addresses. The master node then instructs other hosts to send probe packets with the specific addresses. Depending on where the probe packets are delivered to (or not), a picture of the network internals can be formed. The problem with this method is that special software agents have to be installed on network hosts.

Other efforts worth mentioning are proprietary protocols by network vendors used prior to the standardization of the LLDP. These include the Cisco Discovery Protocol (CDP), Enterasys Networks' Cabletron Discovery Protocol (also CDP), Extreme Networks' Extreme Discovery Protocol (EDP) and Nortel Networks' Nortel Discovery Protocol (NDP).

The use of Ethernet as an access technology, especially in the telecommunication industry, has also led to efforts to add and standardize Ethernet capabilities for Operational, Administration and Maintenance (OAM) management. The main operational issues addressed are discovery, link monitoring, and fault signaling and remote loopback. The added functionality is not aimed specifically at topology discovery in enterprise networks, but could potentially be used.

## V. LIMITATIONS AND ISSUES

Even though the techniques used for network layer topology discovery so far can discover the topology in wide range of case, there are several important cases where these methods may fail to discover the Ethernet network topology.

In practice, network topology can change during the discovery process. Furthermore, AFTs can be stale. Both situations can result in AFTs that are not consists with the actual network topology. Unfortunately, none of the methods published so far in the research community or industry addresses this important issue.

Another limitation with Ethernet topology discovery is the existence of VLANs. In fact, it is very common for Ethernet networks to have VLANs. VLANs are used similarly to subnets but it not necessary and they allow Ethernet networks to spread over large geographical distance.

The main issue with networks that have VLANs is that the network may have cycles and the topology is no longer tree. Breitbart *et al.* described method to discover the topology in the presence of VLANs []. However, since VLANs can spread large geographical areas, and consequently network devices, it is impractical to assume that AFTs will be complete in the presence of VLANs. Figure 3 depicts a typical VLAN and its topological layout.

The third limitation with the current approaches occurs with the existence of wireless and mobile nodes. Wireless and mobile nodes don't follow the classical AFT approach to communicate with other network devices. Consequently, the current methods cannot be reused to infer the topology in the presence of wireless and mobile networks.

## VI. CONCLUSION

The past ten years have seen the rise of a new networking measurement area: the internet topology discovery. Due to its particular structure, the network topology can be understood at various levels. In this article, we focused on the work performed by the research community on the network layer topology, sometimes also called the internet topology.

In this article, we first explained that the internet topology discovery is driven by important questions. For instance, one might want to model the internet in order to reproduce its behavior in a laboratory.

However, although the amount of work performed by the research community is huge, this is not the end of the story. We are starting to see the deployment of large-scale distributed measurement infrastructures made of hundreds or thousands

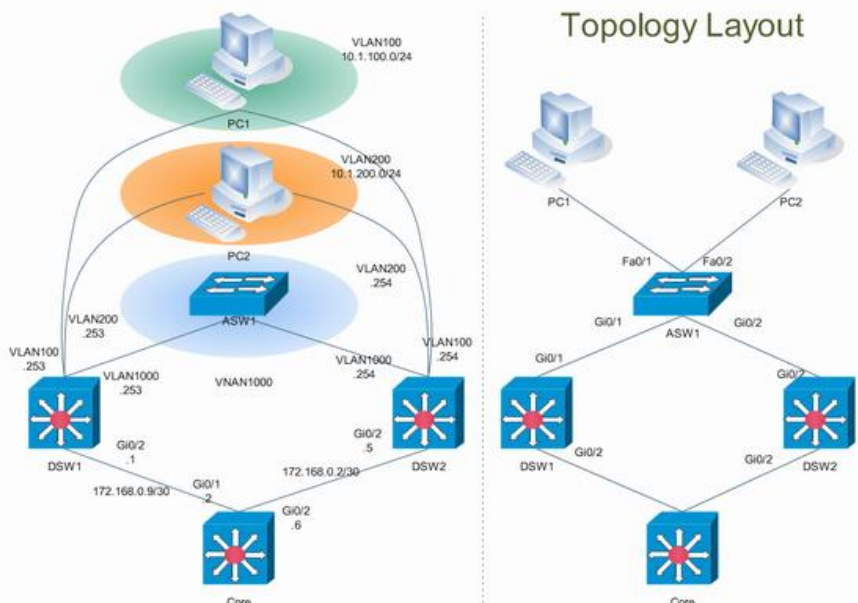

**Figure 3:** A typical VLAN and its topology layout.